\documentclass[aps,pre,twocolumn,superscriptaddress,floatfix,amsmath]{revtex4}
\usepackage{graphicx}
\usepackage{bm}
\usepackage[utf8]{inputenc}
\usepackage{color}
\DeclareUnicodeCharacter{00A0}{~}

\begin{document}
\title{Risk and interaction aversion: screening mechanisms in the Prisoner's Dilemma game}
\author{Gabriel A. Canova}
\author{Jeferson J. Arenzon}

\affiliation{Instituto de Física, Universidade Federal do
Rio Grande do Sul, CP 15051, 91501-970 Porto Alegre RS, Brazil}

\date{\today}

\begin{abstract}
When the interactions between cooperators (C) and defectors (D) can be partially avoided within a population, there may be an overall enhancement of cooperation. One example of such screening mechanism occurs in the presence of risk-averse agents (loners, L) that are neutral towards others, i.e., both L and its opponent, whatever its strategy, receive the same
payoff. Their presence in the Prisoner's Dilemma (PD) game sustains the coexistence of
cooperators and defectors far beyond the level attained in their absence. Another screening mechanism is
a heterogeneous landscape obtained, for example, by site diluting the lattice. In this case, cooperation is enhanced with some fraction of
such inactive, interaction-averse sites. By considering the interplay of both mechanisms, we show that there is an explosive increase in the range of densities, just
above the percolation threshold, where neutrality is prevented and loners become extinct, the behavior reverting to
the pure PD game. Interestingly, this occurs despite defectors being usually abundant in  that region. This has to be
compared with the corresponding loner-free region in the undiluted case that, besides being very small, is dominated
by cooperators.
\end{abstract}
\maketitle

\section{Introduction}
\label{section.introduction}

Risk-sensing is a useful resource~\cite{Miller67,McHo92,ZhBrLo14,HiOlAdHe15} while foraging in changing environments or when deciding the best strategy during a dispute. When participation is not compulsory, withdrawing from a conflict may have several purposes~\cite{Miller67}, like an easy way out, the imposition of some punishment to the opponent or just avoiding a costly situation~\cite{Armstrong84}. 
Evolutionary game theory offers the framework to analyze such conflicting situations as they appear in social dilemmas. In this context, the Prisoner's Dilemma (PD) game sets the paradigm to study how cooperative behavior is seeded and sustained in a  population of cooperators (C) and defectors (D)~\cite{HoSi98,Nowak06a,SzFa07,RoCuSa09c,WaKoJuTa15,PeJoRaWaBoSz17}. Associated with the strategies available for the individuals, there is a payoff that depends on the frequency of other strategies and is directly connected with the fitness, the measure of reproductive success of an individual.    Depending on the strategies chosen by two interacting players, each one receives a payoff that may further affect the temporal evolution of their relative distribution. When both collaborate, each one receives a unitary payoff. Mutual defection  earns both a zero payoff, while a defector obtains $1<b<2$ against a cooperator that, instead, receives nothing. Nonetheless, the fear of being exploited or involved on a risky situation may prompt some agents to avoid playing by the above rules while agreeing on a small payoff $\sigma$ shared with their opponent. With voluntary participation, a risk averse  strategy named loner (L) has been considered both in pairwise~\cite{BaKi95,SzHa02b,SzVu04,WuXuChWa05,HaSz05} and multi-player~\cite{HaMoHoSi02,HaMoHoSi02b,SeKrMi03,SzHa02a,CaTo08} contests (another 
possibility is to use part of the resources to obtain information about the opponent and avoid the conflict accordingly~\cite{SzPe15}).
When pairing is correlated (i.e., on networks~\cite{SzHa02b,SzVu04,WuXuChWa05}, in opposition to fully mixed systems)  because of the corresponding payoffs, the presence of loners diminishes the predatory power of defectors and is indeed an effective strategy against them. However, cyclic dominance~\cite{SzFa07,Frey10,SzMoJiSzRuPe14} emerges once loners can, in turn, be invaded by cooperators, even if not explicitly embedded in the payoff matrix~\cite{HaMoHoSi02,SzHa02b,SeKrMi03,InBiGaKu16}. Such cyclic competition is usually associated with intransitive interactions present in trophic food webs with three or more competing species or strategies (like the Rock-Paper-Scissors game, RPS, and its generalizations~\cite{SzFa07,Frey10,SzMoJiSzRuPe14}), but it is known to appear in several different contexts. Intransitivity decreases, sometimes eliminates, the hierarchy among the trophic levels, thus helping to sustain or enhance coexistence and, as a consequence, biodiversity may persist~\cite{SzFa07,Frey10,SzMoJiSzRuPe14,InBiGaKu16}. Analogously, in the PD game with voluntary participation, because of the emergent cyclic dominance, cooperation can be sustained
even when the temptation to defect ($b$) is large.  In particular, Szabó and Hauert~\cite{SzHa02b} showed that on a square lattice, loners persistently coexist with cooperators and defectors when $b\gtrsim 1.026$, screening their mutual interactions and allowing cooperation to survive up to $b=2$. In the tiny interval $1<b\lesssim 1.026$, loners are suppressed and the standard, two strategies PD game is recovered.

Another screening mechanism known to enhance cooperation is the heterogeneous landscape produced
by randomly diluting the lattice upon which the dynamics unfolds~\cite{VaAr01}. In this case, some patches are permanently voided, offering natural defenses to avoid defection (empty sites may also be interpreted as non-interacting agents, stubborn zealots who receive null payoff and are refractory to the game). 
Some amount of dilution is known to increase the fraction of cooperators within a population~\cite{VaAr01,WaSzPe12a,WaSzPe12b,YaRoWa14,XuZhDuChHu16}, with its peak approaching the random site percolation threshold as the temptation gets closer to the maximum value still allowing cooperation.  The reason is that, as the conditions become harsher for cooperators when $b$ increases, the smaller connectivity provided by the fractal network close to the percolation threshold prevents defectors from disrupting the clusters of Cs.  Once the network is not fully populated, agents may take advantage of the free space and move. In the RPS game, mobility is known to jeopardize the coexistence of the three strategies above a certain threshold~\cite{ReMoFr07}, while for the spatial PD game there are scenarios where it sustains cooperation~\cite{VaSiAr07,VaBrAr14} (despite the inherent difficulties both to be performed and compared with the models, initial experiments with humans~\cite{AnToSa15} had not detected any effects of mobility).
 It is thus important to understand, in this case in which cyclic behavior is not explicitly embedded into the payoff or interaction matrix, whether the behavior displayed in the regular lattice is robust against dilution and mobility. A first step would be the introduction of simple, random mobility on the available space~\cite{VaSiAr07,VaBrAr14}. From that, more complex movement strategies~\cite{Aktipis04,HeYu09,WuHo09,JiWaLaWa10,ChGaCaXu11,ScBe12,WaChWaJiLi13,ZhSuWeXi13,IcSaSaWi13,GeCrFr13,ToAn15,BuLoScHuCh17} may evolve, probably requiring higher cognitive skills. As an application, such simple models of mobile particles help to better understand the  complex behavior of collectively organized robots and other self-propelled particles through disordered environments~\cite{MoDeCaBa16}. Thus, the main question dealt with in this paper is whether and under which conditions the coexistence of all three strategies is possible and the density of cooperators enhanced when the screening mechanisms discussed above are both present and mobility is, eventually, included. Other forms
of disorder have also been considered, for example, in the invasion rates~\cite{SzPe16b,SzPe16c}.

The paper is organized as follows. We summarize the model and give some details of the simulation in Sect.~\ref{section.model}. Then, in Sect.~\ref{section.results} we present results for disordered environments in the absence (Sect.~\ref{section.nodiffusion}) or presence (Sect.~\ref{section.diffusion}) of mobility. Finally, we discuss these results and present our conclusions in Sec.~\ref{section.conclusions}.



\section{Model}
\label{section.model}

In the PD game with the loner strategy, the benefit from withdrawing participation ($\sigma$) is worse than what is received from mutual cooperation, yet better than the punishment for mutual defection, i.e., $0<\sigma<1$.  The payoff matrix is given by 
\begin{align}
\bordermatrix{~ & C & D & L \cr
C & 1 & 0 & \sigma \cr
D & b & 0 & \sigma \cr
L & \sigma & \sigma & \sigma \cr}.
\label{eq.payoff}
\end{align}
The interactions between cooperators and defectors correspond to the weak version of the PD game, leaving $b$, the temptation to defect, as a further parameter. Cyclic dominance is not directly embedded in the payoff matrix~\cite{SzHa02b} since loners, whether interacting with either cooperators or defectors, receive the same payoff. 

We consider a square lattice, with linear size $L$ and periodic boundary conditions. Initially, each site has an equal probability $\rho/3$ of being C, D or L, where $0<\rho\leq 1$ is the overall lattice occupation. In each round, after accumulating the payoffs from the interactions with all its nearest neighbors, player $i$ randomly chooses one of them, say $j$, and, if occupied, tries to copy its strategy with probability  
\begin{equation}
w(S_i\leftarrow S_j) = \frac{1}{1+{\rm e}^{(P_i - P_j)/K}},
\label{eq.prob}
\end{equation}
where $P_i$ and $P_j$ are, respectively, the payoffs obtained by strategies $S_i$ and $S_j$, and $K$ is a (small) noise allowing irrational choices (we take $K=0.1$ hereafter). This updating rule is synchronously applied on all players at each (Monte Carlo) step and keeps $\rho$ unchanged. 
Parallel updating allows us to take full advantage of the GPU computing power whose details are discussed in the Appendix~\ref{ap.gpu} (see also Ref.~\cite{Perc17}). Since the results for $\rho=1$ are qualitatively the same as those obtained in Ref.~\cite{SzHa02b} with a serial dynamics, the model seems robust against the details of the dynamics. Mobility is  also considered, depending on the existence of neighboring empty sites. After the updating process, each individual randomly chooses a nearest neighbor site and, if empty, switches position with it with probability $m$~\cite{VaSiAr07}. We consider the simplest mobility scenario not requiring any particular skills of the agents. In this way, mobility is unbiased (purely random), non assortative and homogeneous (site and strategy independent).

The use of GPUs allows the Monte Carlo simulations to be performed using very large linear system sizes (up to $L=2048$). Averages were taken after the system reached the stationary state
with error bars smaller than the symbol sizes.  
When there is diffusion or the system is
large enough (self averaging), the temporal average seems to be enough.

\section{Results}
\label{section.results}

When a fraction $1-\rho$ of random sites is kept empty throughout the dynamics, the asymptotic fraction of each strategy becomes density dependent~\cite{VaAr01}.
In the next section we discuss the combined effects of the screening mechanisms mentioned in the Introduction on the population of cooperators, defectors and loners.

\subsection{Disordered Environments}
\label{section.nodiffusion}

\begin{figure}[ht!]
\includegraphics[width = 1.\columnwidth]{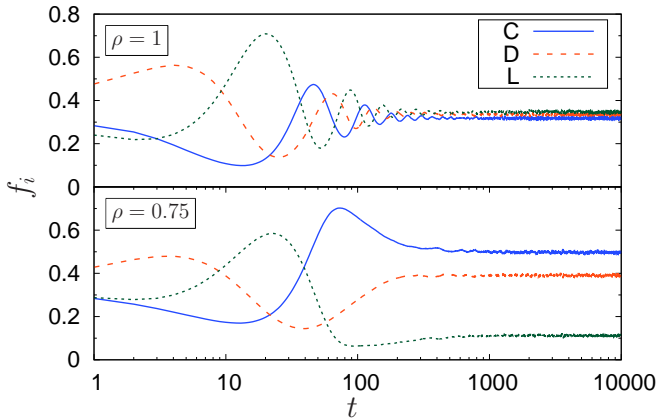}
\caption{Normalized fraction of cooperators (blue, solid line), defectors (red, dashed line) and loners (green, dotted line) as a function of time for $b=1.4$, $\sigma= 0.3$ and two densities: $\rho=1$ (top panel) and $\rho=0.75$ (bottom panel).}
\label{fig.time}
\end{figure}

Fig.~\ref{fig.time} shows the temporal evolution of the fraction of each strategy when $b=1.4$, $\sigma=0.3$ and different densities, $\rho=1$ (top) and 0.75 (bottom panel). The fraction of the strategy $i$ is the ratio between its density $\rho_i$ and $\rho$, $f_i\equiv\rho_i/\rho$. The cyclic dominance induced by the payoff matrix is manifested locally, in each occupied site, as the strategies tend to replace each other as L $\to$ D $\to$ C $\to$ L.
On a global level, although the densities initially oscillate with a macroscopic amplitude,  the  local oscillators are soon driven out of phase, because the interactions are short ranged, and the amplitude decreases while the orbit remains close to the fixed points (some low amplitude fluctuations, due to noise, persist).  The defects introduced in the lattice when $\rho<1$ have an important role in damping the initial oscillations. While for $\rho=1$ the oscillations seem underdamped, for $\rho=0.75$ they are apparently critically damped. In the absence of long range interactions, global synchronization does not occur in this model~\cite{SzVu04}, much akin to related epidemic~\cite{KuAb01} or many species cyclic games~\cite{SzSzIz04,KiLiUmLe05,YiHuWa07,RuAr14}.

\begin{figure}[ht!]
\includegraphics[width = 1.0\columnwidth]{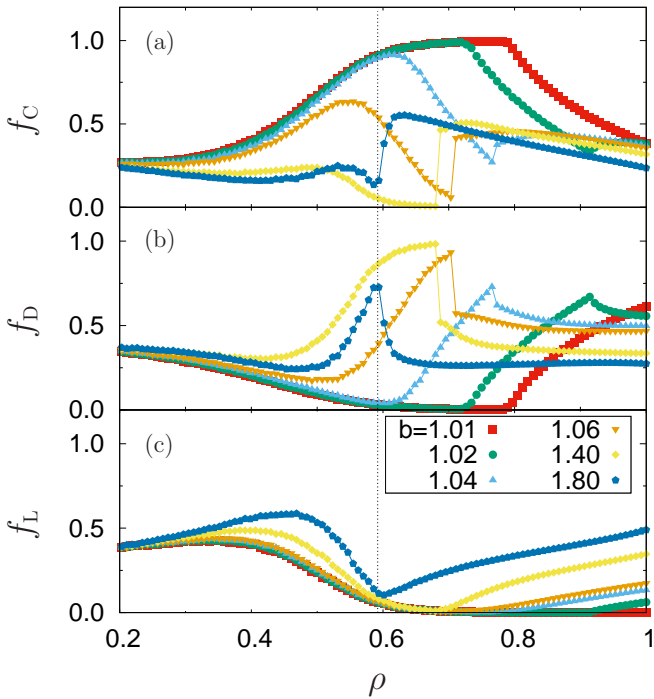}
\caption{Average asymptotic fraction of each strategy as a function of $\rho$ for different values of $b$ and $\sigma=0.3$. From top to bottom: a) cooperators, b) defectors and c) loners. The vertical dotted line locates the random site percolation threshold, $\rho_p\simeq 0.59$ for a square lattice~\cite{StAh94}.}
\label{fig.rho}
\end{figure}

\begin{figure}[ht!]
\includegraphics[width = 1.0\columnwidth]{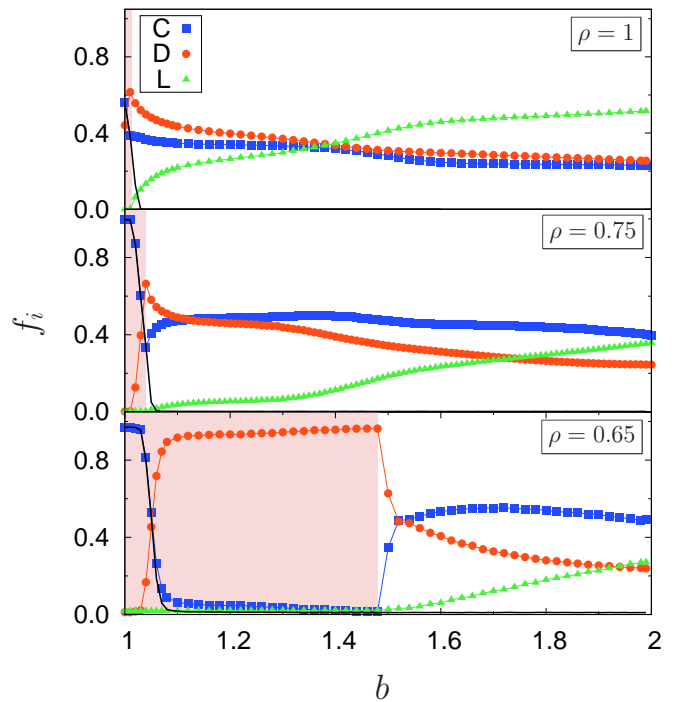}
\caption{Average asymptotic fraction of the three strategies as a function of $b$ for different values of $\rho$ and $\sigma=0.3$. The thick solid lines, for values of $b$ close to 1, show the average fraction of cooperators when there are no loners in the initial state. Inside the shaded region, whose size is much enlarged for $\rho_p<\rho\lesssim 0.7$, the loners present at $t=0$ soon become extinct and the behavior reduces to the pure PD game (the solid, thick line).}
\label{fig.b}
\end{figure}

\begin{figure}[htb!]
\includegraphics[width=1.0\columnwidth]{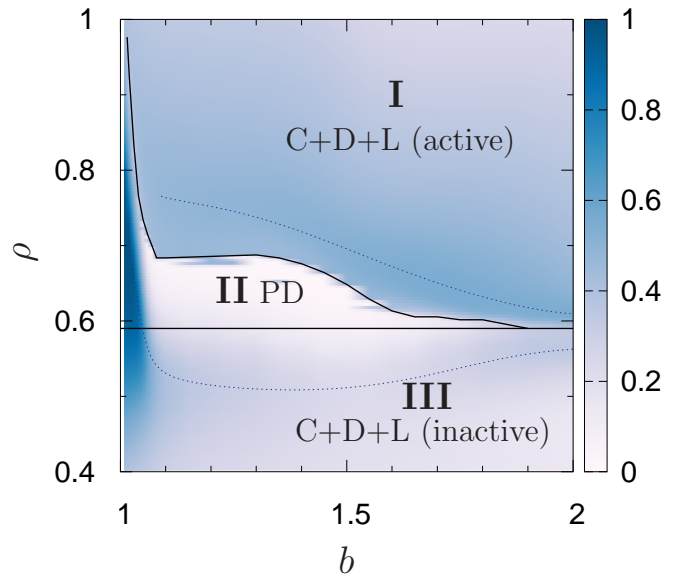}
\caption{Phase diagram with the fraction of cooperators (color code) in the temptation $b$ versus density $\rho$ plane.  The solid lines divide the diagram into three regions with different regimes. In region III, $0\leq\rho\lesssim\rho_p\simeq 0.59$, no activity persists and there is no giant, percolating cluster and the three strategies survive in small, independent clusters. In region I, loners coexist with cooperators and defectors and a finite fraction of agents change strategy at each step (active sites). In region II, instead, loners become extinct and the behavior reduces to the pure PD game. The dotted lines, both above and below $\rho_p$, correspond to the maximum local value of $f_{\scriptscriptstyle\rm C}$.}
\label{fig.rho_s03}
\end{figure}

When there are no loners in the initial state and $b$ is small enough, the fraction of cooperators, $f_{\scriptscriptstyle\rm C}$, has a maximum at an intermediate density $\rho$~\cite{VaAr01}, showing that some amount of dilution enhances cooperation (see the $b=1.01$ curve in the top panel of Fig.~\ref{fig.rho}) by providing a screening mechanism between cooperators and defectors. When $b$ gets close to the point beyond which defectors completely invade the system, the position of the peak seems to approach the random site percolation threshold~\cite{WaSzPe12a}, $\rho_p\simeq 0.59$ for the square lattice~\cite{StAh94} with a dynamics allowing irrational behavior,  Eq.~(\ref{eq.prob}). Cooperators are more easily outperformed  by defectors as $b$ increases and, in order to protect themselves, groups of Cs should decrease the surface of contact with defectors. Such conditions appear close to $\rho_p$ where the underlying, largest cluster is fractal. As discussed in the introduction, loners also play a screening role, preventing cooperators and defectors from playing the game. When loners are initially present and $\rho=1$, we reproduce the results of Szabó and Hauert~\cite{SzHa02b} (with small quantitative differences due to the parallel dynamics). In this case, Fig.~\ref{fig.b} (top), loners disappear for $b\lesssim 1.013$ and within the tiny interval $1<b\lesssim 1.013$, the original PD outcome is recovered (shaded interval). However,   beyond that interval, and up to $b=2$, cooperators survive in the presence of both loners and defectors. In this range, both $f_{\scriptscriptstyle\rm C}$ and $f_{\scriptscriptstyle\rm D}$ decrease while $f_{\scriptscriptstyle\rm L}$ increases. In particular, as $b$ approaches the critical value from above~\cite{SzHa02b}, $f_{\scriptscriptstyle\rm L}$ vanishes with an exponent $\beta\simeq 0.58$, very close to the directed percolation (DP) universality class~\cite{Hinrichsen00} (interestingly, for $\rho$ not far from unity we still find exponents compatible with the DP universality class, although this is not expected in the presence of quenched disorder). 
Dilution dramatically changes  the value of $b$ above which loners are able to survive in the population. Down to $\rho\simeq 0.7$, the effect is small (compare, for example, the size of the shaded regions in Fig.~\ref{fig.b}).  Then, there is an explosive  
increase and the loner-free region becomes very large (e.g., up to $b\simeq 1.5$ for $\rho=0.65$, bottom panel of Fig.~\ref{fig.b}). In all three panels, below this value (shaded region), loners do not persist and, interestingly,
the asymptotic fraction of cooperators is almost identical to the one obtained with loners absent already in the initial state (solid thick lines), i.e., the initial presence of loners does not change the asymptotic density, despite the strong transient oscillations (Fig.~\ref{fig.time}). In the
region indicated by II in the phase diagram Fig.~\ref{fig.rho_s03}, the behavior is thus identical to the pure, loner-free PD game. In other words, on a diluted lattice, there is a rather broad range of densities, $\rho_p<\rho\lesssim 0.7$, right above the percolation threshold (i.e., a landscape still connected) where neutral strategies are inhibited and become extinct, the behavior reducing to the standard PD game. The behavior of $f_{\scriptscriptstyle\rm C}$ becomes non-monotonic (see, for example, the bottom panel of Fig.~\ref{fig.b}, for which there is even  a local optimum around $b\simeq 1.7$). Interestingly, for $\rho_p<\rho\lesssim 0.7$, cooperators reappear, after being suppressed in region II, when conditions become more favorable to defectors, i.e., as $b$ increases.

\begin{figure}[ht!]
\includegraphics[width = 0.9\columnwidth]{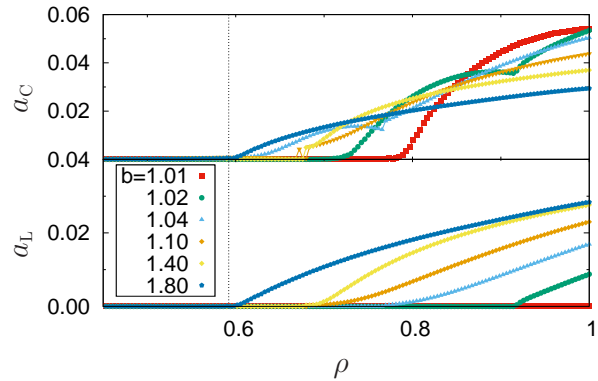}
\caption{Fraction of active cooperators (top) and loners (bottom). Defectors (not shown) behave in a similar way to cooperators. Notice that below the random site percolation threshold ($\rho_p\simeq 0.59$, vertical dotted line~\cite{StAh94}), there are no active strategies whatsoever.}
\label{fig.active}
\end{figure}

More specifically, the behavior of $f_i$ as a function of $\rho$ is shown in Fig.~\ref{fig.rho}.  For $\rho\to 0$ all occupied sites are isolated and keep the strategy initially assigned to them, i.e., $f_i=1/3$, $\forall i$. For values of $\rho$ well below the percolation threshold, the system is divided into small, independent clusters whose fate depends essentially on their initial composition~\cite{VaAr01}: both $f_{\scriptscriptstyle\rm C}$ and $f_{\scriptscriptstyle\rm D}$ initially decrease as $\rho$ increases, while loners inside these small clusters are more successful and $f_{\scriptscriptstyle\rm L}$ increases. In this region, labeled III in Fig.~\ref{fig.rho_s03} and limited from above by $\rho_p$, the system attains an inactive, static pattern without sites that change strategy, as shown in Fig.~\ref{fig.active}, with the possible exception of a few blinkers.
Upon further increasing $\rho$, a strong dependence on $b$ appears, with interesting characteristics close to and above the percolation threshold $\rho_p$.
For $\rho\geq\rho_p$, loners survive on the percolating cluster only in phase I, while they are strongly suppressed in phase II (where they may survive in some of the small, isolated clusters coexisting with the percolating one). 
Once inside phase I, loners also become active, $f_{\scriptscriptstyle\rm L}$ resumes growing with its density being correlated with its activity. 
Above the value of $b$ where phase II disappears, there remains a minimum of $f_{\scriptscriptstyle\rm L}$ at $\rho_p$ (see, e.g., the case $b=1.8$ in Fig.~\ref{fig.rho}c), indicating that loners thrive better in a compact environment, the fractal nature of the percolating cluster at $\rho_p$ not providing enough support for them. 

While dilution may enhance cooperation in both two and three strategies games, the precise dependence of the optimal density on $\rho$ is now more complex. 
Indeed, as $b$ increases from 1, the peak of cooperators shrinks and move to the left, toward $\rho_p$ (see Fig.~\ref{fig.rho}a). 
 However, differently from the case without loners~\cite{WaSzPe12a}, the peak does not monotonously converge to $\rho_p$ but, instead, continues into the $\rho<\rho_p$ region while also decreasing in height (see, e.g., the $b=1.06$ case in Fig.~\ref{fig.rho}a). In addition, a second peak  develops inside phase I and also  approaches $\rho_p$ (from above) as $b\to 2$. Both peaks are shown as dotted lines in  Fig.~\ref{fig.rho_s03}. Loners thus affect the subcritical clusters below $\rho_p$, reducing the presence of defectors and allowing cooperators control those large, but finite domains in that region.


Finally, another interesting feature can be observed  for large values of $b$ (e.g., $b=1.8$ in Fig.~\ref{fig.rho}b). Almost everywhere above $\rho_p$, the density of defectors
is essentially flat 
The frequency of loners increases as the lattice becomes more populated, while the frequency of their predators (the cooperators) decreases, what is reminiscent of the survival of the weakest effect~\cite{FrAb01}. 

\subsection{Mobility}
\label{section.diffusion}

\begin{figure}[ht!]
\includegraphics[width = 1.\columnwidth]{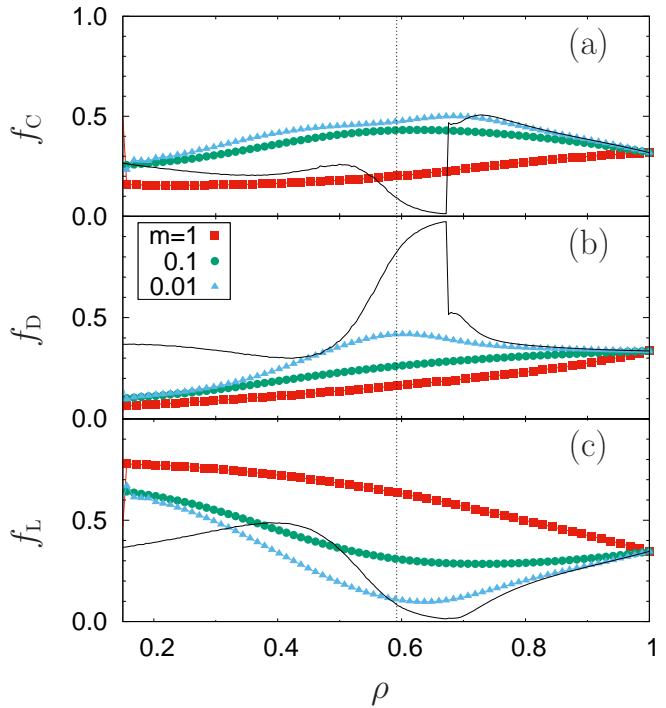}
\caption{Fraction of cooperators (a), defectors (b) and loners (c) as a function of $\rho$ for $b=1.4$, $\sigma=0.3$ and several values of the mobility $m$. The solid lines, for comparison, are for $m=0$, Fig.~\ref{fig.rho}. The behavior is qualitatively the same for others values of $b$ as long as $m>0$.}
\label{fig.painel.mob1}
\end{figure}

\begin{figure}[ht!]
\includegraphics[width = 1.\columnwidth]{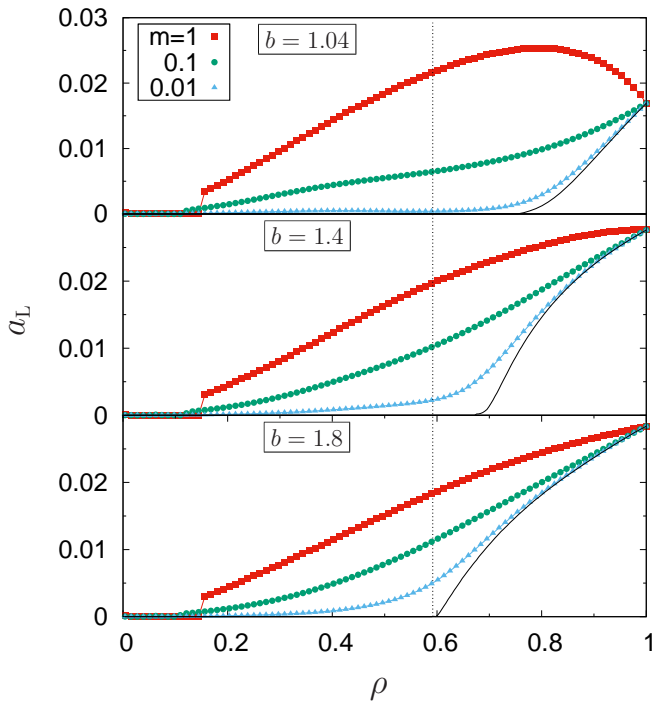}
\caption{Fraction of active loners associated to Fig~\ref{fig.painel.mob1}. The transition between phases I and II disappears.}
\label{fig.painel.ativos.m}
\end{figure}

Mobility does not need to be uncorrelated with strategy, payoff, environment, expectations, etc, although the simplest case of random diffusion settles the background against which to compare other possible scenarios. In particular, it may be even independent of the occupation $\rho$ of the lattice if implemented through, for example, the swapping of two neighboring strategies. Nonetheless, we here explicitly consider a rule allowing only diffusive steps toward empty sites in order to better understand the role of the disordered environment on the behavior of the model. Results for non-assortative, homogeneous random diffusion are shown in Fig.~\ref{fig.painel.mob1} for $b=1.4$, $\sigma=0.3$ and several values of the mobility $m$ (including the highly viscous, $m=0$ case, solid lines, for comparison). Although the behavior for $m=0$ strongly depends on the value of $b$, once mobility is
included it becomes more homogeneous, with small quantitative differences. 
When the mobility is small, cooperation levels increase in the regions previously occupied by phases II and III, while remaining almost the same in phase I. In almost all cases,  $f_{\scriptscriptstyle\rm C}$ and $f_{\scriptscriptstyle\rm D}$ decrease as $m$ increases. Interestingly, loners seem to benefit from high levels of mobility. In particular $f_{\scriptscriptstyle\rm L}$ is an almost monotonously increasing function of $m$ for all densities (except for extremely low $m$). 
Fig.~\ref{fig.painel.ativos.m} shows the average fraction of active loners for different values of $b$.
For $m=0$, phase I was characterized by the presence of all three strategies being active, while in II and III loners were either not present or inactive. For $m>0$,  there is no longer such distinction and the active coexistence extends down to very low densities, even below the percolation threshold (for very low densities data is not shown because it takes takes too long to reach the steady state). 
Diffusion seems to be, therefore, beneficial to the maintenance of coexistence.

Interestingly, for low densities (and $m>0$)  
there is a sea of loners where small and isolated groups of cooperators concentrate on larger metadomains that are swapped by invading waves of defectors. The absence of the reassuring effects of large groups of cooperators in this regime is the associated risky condition that benefits the loner strategy. 
In this scenario of small clusters, changing the value of $m$ only slightly change the average frequency of each specie. 
For very low densities (in the range $\rho\simeq 0.15$ and below) 
it becomes very easy to be 
absorbed after an extinction due to demographic stochasticity of finite systems.

\section{Conclusions}
\label{section.conclusions}

Game theory gained its evolutionary version once the more effective scale of selection, the individual, was clearly identified~\cite{Hammerstein98} (it is possible, nonetheless, to argue for higher levels of selection~\cite{Dugatkin98,Wilson98}). Differently from the applications of game theory in economy, the decision process is now ruled by natural selection based on the individual behaviors (strategies) of all agents. Specifically, selection is frequency-dependent and acts on the phenotype characteristics, often neglecting any underlying genetic mechanisms. The problem of the persistence of cooperation has been challenging science for decades~\cite{Pennisi05} and many mechanisms have been proposed and tested experimental and numerically. 
When participation is voluntary, agents that are risk-averse may avoid the competition and deter exploitation by assuming a neutral (loner) strategy whose payoff is independent of the strategy of the opponent. 
Such neutrality, amidst a population, acts as a screening mechanism that decreases the average strength of the actual PD interactions.  Further screening is obtained by considering disordered environments through a diluted lattice, where empty sites may be interpreted as interaction-averse agents. Once space is available, avoiding the game may also be achieved by moving away from the opponent.  
We here combined such mechanisms, and study the effects of disordered environments (random dilution) and mobility on the PD game with loners, cooperators and defectors. The corresponding payoff matrix, Eq.~(\ref{eq.payoff}), induces a cyclic behavior, similar to the Rock-Scissors-Paper game, where loners prey on defectors that prey on cooperators that, in turn, prey on loners. Such mechanism is known to reduce the hierarchy among the strategies, leading to coexistence. 
Thus, besides the formation of groups as the survival mechanism of cooperators in a spatial setup, in the presence of loners, cyclic competition is a further resource enhancing diversity and sustaining cooperation.
Moreover, in the original PD game, dilution along with an increasing temptation to defect pushes the optimal conditions for cooperators toward the percolation threshold because the existence of a giant, non-compact cluster provides enough support for cooperators to thrive while its filamentous nature prevents an efficient exploitation by defectors. 
In the undiluted case, $\rho=1$, there is a tiny interval of $b$ that suppresses loners but outside it, they sustain cooperators in the whole interval of $b$.
The main result shown here is that, in the presence of dilution (on a broad range close to $\rho_p$), there is an explosive increase in the region (II) excluding neutral strategies. Importantly, most of phase II is defector rich. As a consequence, loners become extinct in that region not because preys are absent or reduced in number (as the result for $\rho=1$ might have implied) but due to a different mechanism, directly related to the diluted lattice. 
We verified that this explosive expansion of the pure PD phase  also appears in other geometries (random graph, honeycomb and triangular lattices).

The behavior in region II originates from how the interfaces between strategies respond differently to the pinning effects of dilution. While the velocity of a flat interface separating loners from cooperators or defectors does not depend, obviously, on $b$ and is barely affected by the empty sites, an interface between Cs and Ds increases its velocity with $b$ and becomes pinned if the dilution is high (or $K=0$). The smaller $b$ is, the larger is the number of neighboring cooperators that the defector needs, and the behavior of the defector invasion front is reminiscent of the partially directed percolation problem~\cite{MaVa85}, albeit the situation here is more complicated due to two reasons. First, because of the random initial state, each interface is directed differently. Second, for an invading front, only the sites with the prey strategy are relevant, the others (including the empty sites) only disrupt the movement. Because of that, the transition to phase II occurs at a value slightly larger than the partially directed percolation threshold, $\rho_{\scriptscriptstyle\rm pDP}\simeq 0.64$~\cite{MaVa85}.  Together, just above the percolation threshold (still high dilution) and $b$ not so large, i.e., region II, these effects make the domain interfaces between cooperators and defectors almost immobile. The interfaces between loners and defectors, however, are still mobile and survive inside the clusters of the latter as long as they are reasonably large and interconnected. But in region II these clusters become small and disconnected, leading to the extinction of loners.

Under more realistic conditions, dilution is not expected to be spatially uncorrelated. In those cases, correlated~\cite{CoFi16} rather than random percolation should be more relevant, along with mobility being a further essential ingredient for agents to self-organize in patchy communities.
Movement through the available space was considered here with agents jumping to neighboring empty sites with probability $m$, independently of their strategy, i.e., homogeneous, purely random, non assortative diffusion.
Nonetheless, being another mechanism to avoid risk, it is possible to have heterogeneous, somewhat assortative mobility~\cite{ChLiDaZhYa10}, with loners having a different probability, $m_{\scriptscriptstyle\rm L}$, than the other agents. In one extreme case, loners may avoid risk by drifting away from their non-loners, immobile neighbors ($m=0$, $m_{\scriptscriptstyle\rm L}\neq 0$), what can be interpreted as fleeing from risky situations without helping others~\cite{MoTr16} or, alternatively, as a punishment for not complying with the social norm and, consequently, being expelled. In the other extreme,  
$m\neq 0$ and $m_{\scriptscriptstyle\rm L}=0$, loners do not care much about their neighborhood and stay put while the other strategies may move. Obviously, intermediate cases are possible as well.  A similarly heterogeneous mobility among cooperators, defectors and loners, but hybrid between random and driven, was also considered in Ref.~\cite{ZhXuShQi13} in the context of the Public Goods Game.
As originally discussed in Ref.~\cite{VaSiAr07}, once agents are able to diffuse they may evolve more sophisticated mobility forms, no longer purely random but, perhaps, with strategy, payoff, neighborhood, history, etc, dependence. 
It would be also interesting to include other coexisting strategies, like the Tit-for-Tat or a more tolerant version~\cite{SzPe16a} of the pure loner considered here. One still open question is how similar this system is to cyclic competition models that go beyond the RSP game with more complex, multi-looped food webs~\cite{SzFa07,SzMoJiSzRuPe14} (where another coexistence mechanism, defensive alliances, becomes important) and whether the pure PD phase obtained here in the presence of dilution is a consequence of the neutrality of loners or another mechanism. 


\begin{acknowledgments}
We thank H.C.M. Fernandes, A.R. de la Rocha, and M.H. Vainstein for discussions and collaboration at the early stages of this project.
Research partially supported by the Brazilian agencies CNPq, CAPES and FAPERGS. JJA also
acknowledges partial support from the INCT-Sistemas Complexos.
\end{acknowledgments}

\appendix
\section{GPU computing}
\label{ap.gpu}

The Monte Carlo simulations were implemented on a square lattice with periodic boundary conditions. The strategy on each lattice site is represented by the variable $s_i=0,\ldots,3$ corresponding, respectively, to an empty site, cooperator (C), defector (D) and loner (L). 
For a given total density $\rho$, to each site  a random strategy is initially assigned with probability $\rho/3$. 
The combat, the first step of each round, where all agents interact with their nearest neighbors, is easily done in parallel since every player gets its payoff independently from each other.

During the selection step, each individual randomly chooses one of its neighbors and changes its strategy accordingly with the dynamic rule Eq.~(\ref{eq.prob}). First, to avoid memory overwrite, the state ${\bm s}$ is saved and then each site runs at parallel (attempts to optimize the memory read efficiency by copying blocks of the strategies to the shared memory have been fruitless).    
It is worth mentioning that the computationally expensive evaluation of the exponential in Eq.~(\ref{eq.prob}) can be avoided by noticing that $e^{(M_i - M_j)/K}=e^{M_i/K}/e^{M_j/K}$, then we can write $e^{M_i/K}=e^{M_E/K}.e^{M_W/K}.e^{M_N/K}.e^{M_S/K}$, where $M_E,M_W,\ldots$ are the pairwise payoffs from the $i$ neighboring sites. Therefore, we can pre-evaluate those pairwise interactions and save them into an array $P(s_i,s_j)$ in the constant memory. Finally, one can write 
\begin{equation}
e^{(M_i - M_j)/K}=\frac{\displaystyle\Pi_{k=0}^4 P(s_i,s_k)}{\displaystyle\Pi_{l=0}^4 P(s_j,s_l)}, 
\end{equation}
where $k$ e $l$ run over the $i$ and $j$ neighborhood, respectively.

\begin{figure}[hbt!]
\includegraphics[width = 0.3\columnwidth]{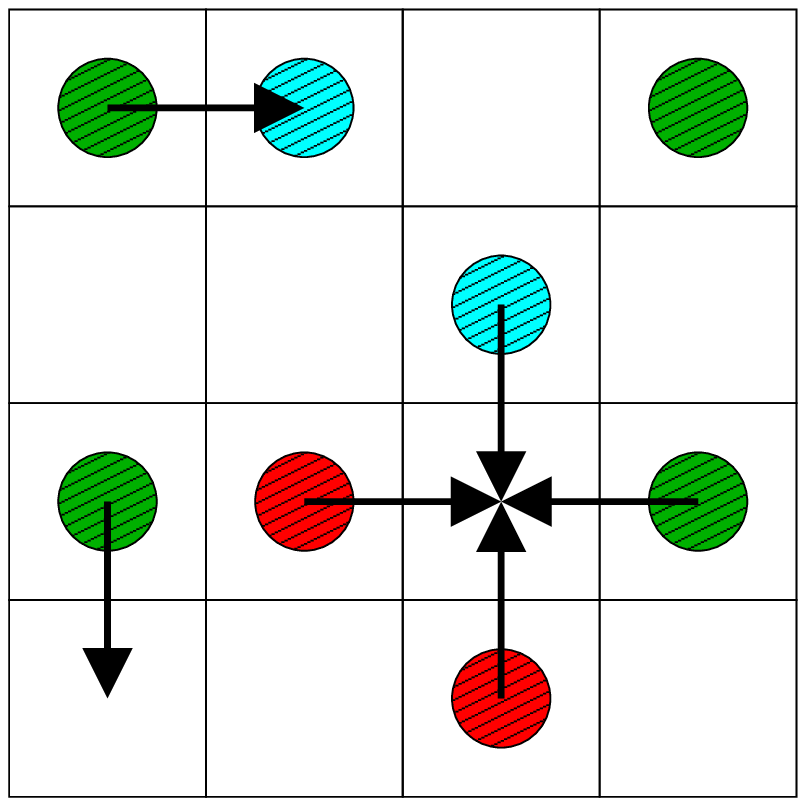}\hspace{1cm}
\includegraphics[width = 0.3\columnwidth]{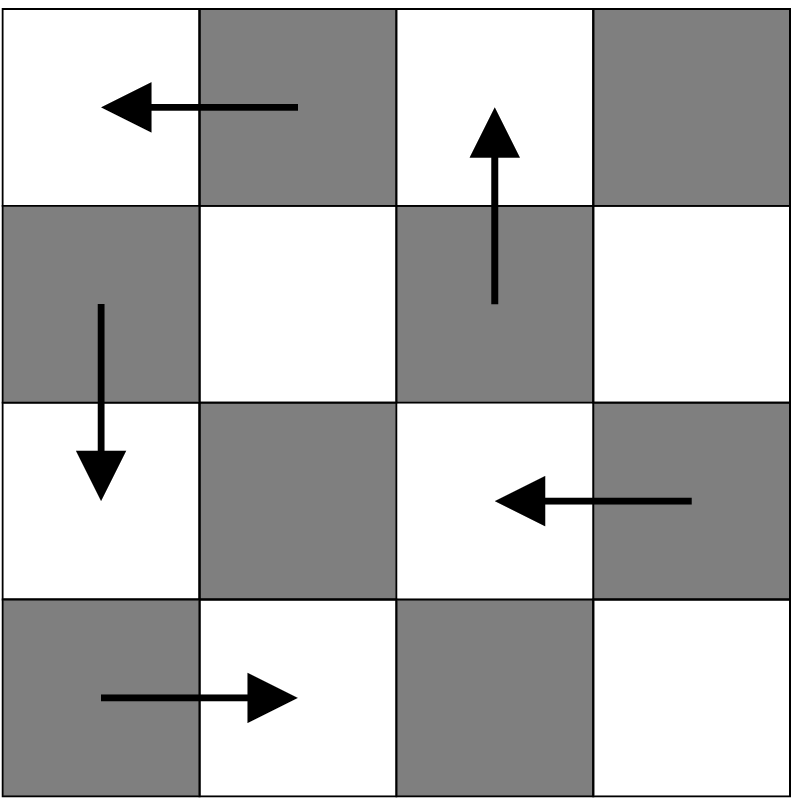}\\
\caption{(Left) With probability $m$, each player chooses a destination to migrate. (Right) The white (black) empty sites are occupied by   players coming  from the black (white) sites.}
\label{chess}
\end{figure}

To proceed with the parallel diffusion, three consecutive steps are necessary. First, each player chooses, with probability $m$, one of its neighboring sites (empty or not) as a possible destination, Fig.~\ref{chess}a. Then, every site on the lattice, with the information of those neighbors that are attempting to occupy it, randomly chooses one of the contenders (when there is more than one). At last, the lattice is divided into sublattices, the chessboard pattern in Fig.~\ref{chess}b. As a first movement, the white empty sites let the chosen individuals standing on the black ones migrate to them. In the next turn, the black sites repeat the process.

\bibliographystyle{apsrev4-1}

%

\end{document}